\documentclass[12pt]{article}
\title{Duality in the BTZ Black Hole and the Statistical Entropy}
\author{Leopoldo A. Pando Zayas\\
Department of Mathematics\\
University of Wisconsin--Green Bay, WI 54311\\
{\tt pandol@gbms01.uwgb.edu}}
\date{}

\def\a{\alpha}
\def\b{\beta}
\def\o{\omega}

\def\t{\tau}
\def\te{\theta}
\def\p{\phi}

\def\r{\rho}
\def\hhr{\hat{r}}

\def\hht{\hat{t}}
\def\m{\mu}
\def\n{\nu}
\def\na{\nabla}

\def\wg{\tilde{g}}
\def\wB{\tilde{B}}
\def\hht{\hat{t}}
\def\hhx{\hat{x}}
\def\hhr{\hat{r}}
\def\ts{\tilde{s}}
\def\cq{{\cal Q}}
\def\cm{{\cal M}}

\begin{document}

\maketitle

\begin{abstract}
In Strominger's proposal for the computation of the statistical entropy of
black holes based on the asymptotic symmetry analysis of Brown
and Henneaux a fundamental role is played by the asymptotic conditions that
the considered metric must satisfy at infinity. Here it is shown that
$T$-duality does not preserve such conditions. This observation is used to
discuss a possible reformulation of the  proposal.
\end{abstract}


The search for the ultimate foundation for the black hole statistical entropy
is now the subject of many investigations. Due to the fact that string theory
has been unable to solve this problem in the generic case, that is, including
nonsupersymmetric black holes  (see \cite{peet} for an extensive review), new
directions are being explored. A very strong approach has been proposed by
Strominger \cite{st} where the problem is reduced, in the case of black
holes with near horizon geometry with an $AdS_3$ factor, to the counting of
microstates for the Ba\~nados-Teitelboim-Zanelli black hole \cite{btz}.
Strominger's proposal has been tested in a variety of black holes and
exact numerical agreement has been found in all cases
\cite{bir,bl,teo,plb3,ka,ca,prd,bbg,cl,cl1}.  The formulation of the proposal
is not quite definite yet and further investigation has been
directed to the validity of the statistical counting of the microstates of
the BTZ black hole itself \cite{ms,ma,bbcho,le,cawh,boer,se,bbg1,my} in some
cases in the scope of the AdS/CFT correspondence \cite{largen}.

A very exciting feature of this approach is that it does not rely on any
specific quantum theory of gravity. Along these lines of reasoning it seems
that finding the minimal set of conditions necessary for performing the count
of microstates in the BTZ black hole is not merely a very interesting problem
{\it per se} but it is a problem that will certainly shed some light on the
common features that any consistent quantum theory of gravity must share.
The purpose of this paper is to take a step in this direction by studying the
role of the geometric conditions used in \cite{st}.

The initial motivation for this work arose from the following puzzling fact:
Strominger's proposal focuses in the near horizon geometry asserting that the
statistical entropy of any black hole whose near horizon geometry contains an
$AdS_3$ factor can be computed using the statistical counting of
microstates of the BTZ black hole. At the same time the counting of
microstates proposed is based in the asymptotic form of the metric at
infinity \cite{brhe}\footnote{The fact that the BTZ entropy can be obtained
from the results of Brown and Henneaux was obtained independently in
\cite{bss}.}.  In most of the analyzed cases the near horizon region is at
$r\to 0$ while the conformal field theory lives at $r\to \infty$. This
clearly suggests a kind of $r\to 1/r$ relation which is nothing but the
cornerstone of $T$-duality.  Some arguments have been presented to fill this
gap, most notably in \cite{ba} and recently in \cite{ma}. In this paper we
present an argument that shows the near horizon geometry as being more
fundamental than the asymptotic conditions to the formulation of \cite{st}.
The argument is based on the fact that $T$-duality does not preserve the
asymptotic conditions and thus any argument based on them is not compatible
with string theory.


The low energy string effective action that we will consider in this paper is
$(\a'=1)$

\begin{equation}
\frac{1}{16\pi G}\int d^3x\sqrt{-g}e^{-2\p}\left(R
+4(\na\p)^2-\frac{1}{12}H^2+\frac{4}{l^2}\right).
\end{equation}
The conformal field theory origin of the last term is a central charge
deficit but from the gravitational point of view it is a cosmological
constant. Using dualities it is possible to consider not only the heterotic
case.  Given a solution to the equations of motion implied by this action
$(g_{\m\n},B_{\m\n},\p)$ that is independent of one coordinate, say $\te$,
there exists another solution to these equations of the form \cite{bu}

\begin{eqnarray}
\label{dual}
\wg_{\te\te}&=&1/g_{\te\te}, \qquad \wg_{\te\a}=\frac{B_{\te\a}}{g_{\te\te}}, \qquad
\wB_{\te\a}=\frac{g_{\te\a}}{g_{\te\te}},
\qquad \tilde{\p}=\p-\frac{1}{2}\ln g_{\te\te},\nonumber \\
\wg_{\a\b}&=&g_{\a\b}
-\frac{g_{\te\a}g_{\te\b}-B_{\te\a}B_{\te\b}}{g_{\te\te}}, \quad
\wB_{\a\b}=B_{\a\b}
+\frac{g_{\te\a}B_{\te\b}+B_{\a\te}g_{\te\b}}{g_{\te\te}}. \nonumber \\
\end{eqnarray}

Central to Strominger's proposal is a result of Brown and Henneaux
\cite{brhe} stating that in any consistent gravity theory with metric
with the following asymptotic behavior \footnote{This behavior is inspired by
$AdS_3$}

\begin{eqnarray}
\label{asy}
g_{tt}&=&-\frac{r^2}{l^2}+{\cal O}(1), \quad g_{t\p}={\cal O}(1), \nonumber \\
g_{\p\p}&=&r^2+{\cal O}(1), \qquad g_{r\p}={\cal O}(\frac{1}{r^3}),\nonumber\\
g_{rr}&=&\frac{l^2}{r^2}+{\cal O}(\frac{1}{r^4}),\quad
g_{tr}={\cal O}(\frac{1}{r^3}),
\end{eqnarray}
there is an underlying conformal field theory associated with the asymptotic
symmetry group. To make the $T$-duality analysis relevant to the work of
Brown and Henneaux one has to consider the Einstein metric
$g_{\m\n}^E=e^{-4\p}g_{\m\n}^S$ which under the duality (\ref{dual})
transforms as

\begin{eqnarray}
\wg^E_{\te\te}&=&e^{-4\tilde{\p}}\wg_{\te\te}=g_{\te\te}^E \nonumber \\
\wg_{rr}^E&=&g_{rr}^Eg_{\te\te}^2-g_{\te\te}^E\left((g_{r\te})^2-(B_{\te
r})^2\right)
\end{eqnarray}

This shows that in the general case the  asymptotic conditions (\ref{asy})
are not preserved under the duality transformation (\ref{dual}), and
therefore, the computation of the statistical entropy of 3D string black holes
using the Virasoro algebra associated with diffeomorphism invariance of
metrics with asymptotic behavior (\ref{asy}) needs further investigation
within string theory.

A very important example is the BTZ black hole which is a solution of 3D
string theory \cite{ho}  with the following background

\begin{eqnarray}
\label{btzs}
ds^2_{BTZ}&=& -N^2dt^2+r^2(d\te+N_{\te}dt)^2
+N^{-2}dr^2, \nonumber \\
N^2&=& \frac{r^2}{l^2}+\frac{16J^2G^2}{r^2}-8GM, \quad
N_{\te}=-4 \frac{JG}{r^2}, \nonumber \\
B_{\te t}&=&\frac{r^2}{l^2}, \quad \p=0.
\end{eqnarray}
This background satisfies the asymptotic conditions (\ref{asy}). Dualizing
over the $\te$ coordinate yields

\begin{eqnarray}
d\tilde{s}^2&=&-\left(\frac{16J^2G^2}{r^2}-8GM\right)dt^2+\frac{2}{l}d\te dt+
\frac{1}{r^2}d\te^2+\left(\frac{r^2}{l^2}-8GM
+\frac{16J^2G^2}{r^2}\right)^{-1}dr^2 \nonumber \\
B_{\te\t}&=&-\frac{4JG}{r^2}, \quad \p=-\ln r.
\end{eqnarray}

The asymptotic behavior of this metric is not of the form  required in
(\ref{asy}), in particular
\begin{equation}
\wg_{rr}^E=r^4 \left(\frac{r^2}{l^2}-8GM+\frac{16J^2G^2}{r^2}\right)^{-1}
\sim r^2.
\end{equation}
This result logically prevents us from studying this background along the
lines of \cite{st}.

Changing to the following coordinates \cite{ho1}
\begin{equation}
\label{ch}
t=\frac{l(\hhx-\hht)}{(r_+^2-r_-^2)^{1/2}}, \quad
\te=\frac{r_+^2\hht-r_-^2\hhx}{(r_+^2-r_-^2)^{1/2}}, \quad r^2=l\hhr,
\end{equation}
where $8GM=\frac{r_+^2+r_-^2}{l^2}, \quad 8GJ=\frac{2r_+r_-}{l}$,
the solution becomes

\begin{eqnarray}
d\tilde{s}^2&=&-(1-\frac{{\cal M}}{\hhr})d\hht^2+(1-\frac{{\cal Q}}{{\cal
M}\hhr})d\hhx^2
+(1-\frac{{\cal M}}{\hhr})^{-1}(1-\frac{{\cal Q}}{{\cal M}\hhr})^{-1}
\frac{l^2d\hhr^2}{4\hhr^2} \nonumber \\
\p&=&-\frac{1}{2}\ln\hhr l, \quad B_{\hhx\hht}={\cal Q}{\hhr}
\end{eqnarray}
where ${\cal M}=r_+^2/l$ and ${\cal Q}=4GJ$. This solution is precisely the
black string of \cite{hh}. It is important to notice that although the
duality transformation changes the asymptotic behavior from $AdS_3$ to flat,
the entropy of the black string equals exactly that of the BTZ black hole
\cite{ho1}).

One question that in our opinion is at the core of the problem
here is that although the black string is asymptotically flat, its near
horizon geometry is of the $AdS_3$ type. This argument follows essentially in
the same lines as that for the magnetic solution considered in \cite{plb3}. To
visualize that the near horizon geometry is of $AdS_3$ type it is convenient
to change to the following radial coordinate

\begin{equation}
\o^2=1-\frac{{\cal M}}{{\hhr}}.
\end{equation}
The Einstein metric becomes
\begin{eqnarray}
d\ts^2&=&-\left(\o^2\frac{{\cal Q}^2}{{\cal M}^2}
-(1-\frac{{\cal Q}^2}{{\cal M}^2})\right) \frac{{\cal
M}^2l^2}{(1-\o^2)^2}d\hhx^2 +\frac{\o^2{\cal M}^2l}{(1-\o^2)^2}d\hht^2
\nonumber \\
&+&\left(\o^2\frac{{\cal Q}^2}{{\cal M}^2} -(1-\frac{{\cal
Q}^2}{{\cal M}^2})\right)^{-1}\frac{l^4{\cal M}^2}{(1-\o^2)^4}d\o^2.
\end{eqnarray}

Taking $\o\to i\o$ and considering $\hht$ compact and $\hhx$ noncompact
\footnote{See \cite{plb3} for a proof that this transformation is harmless
for $AdS_3$.} and changing to\footnote{This change is motivated by
the change (\ref{ch}),  see \cite{ho1} for a detailed discussion.}
\begin{equation}
\label{ch1}
\r=\o \frac{{\cal Q}l}{(r_+^2-r_-^2)^{1/2}}, \quad
\tau=\hhx \frac{{\cal M}^2l^2(r_+^2-r_-^2)^{1/2}}{{\cal Q}}, \quad
\te=\hht\frac{{\cal M}}{{\cal Q}}(\frac{r_+^2-r_-^2}{l})^{1/2}.
\end{equation}
In the  near horizon limit $\o \to 0 \quad (\hhr\to {\cal M})$  and the near
extremal limit $(r_+,r_-\gg r_+-r_-)$\footnote{The analysis of the extremal
case is more subtle and has been developed in \cite{ka}. Note that in the
extremal case the change (\ref{ch}) and (\ref{ch1}) are not defined.} the
metric can be written as
\begin{equation}
d\ts^2=-\left(\frac{\r^2\cq^2}{\cm^4l^4}-\frac{A^2}{\cm^2l^4}(1
-\frac{\cq^2}{\cm^2})\right)d\t^2 +\r^2 d\te^2+
\left(\frac{\r^2\cq^2}{\cm^4l^4}-\frac{A^2}{\cm^2l^4}(1
-\frac{\cq^2}{\cm^2})\right)^{-1}d\r^2,
\end{equation}
with $A=\cq l/\sqrt{r_+^2-r_-^2}$. This is a nonrotating BTZ black hole with
entropy exactly equal to the starting seed
solution (\ref{btzs})
\begin{equation}
S=\frac{2\pi r_+}{4G}.
\end{equation}
Thus by studying only the near-horizon behavior of the metric one can retrive
the entropy.

It is worth noting that $T$-Duality is intrinsically a string theory symmetry
and therefore the arguments of this paper do not apply to other gravity
theories. The fact that $T$-Duality trades momentum modes (associated with
any field theory) for winding modes (inherently stringy degrees of freedom)
is arguably the fundamental explanation of the mismatch found in
this paper\footnote{I thank Robert Myers raising this point}. This point of
view proved very useful in a similar case treated in \cite{duff}.

Putting together some results on the near horizon approach to the statistical
entropy and using very elementary mathematical vocabulary the situation seems
to be the following. String theory embedding of a solution is a sufficient
condition for the counting of microstates using Strominger's proposal. The
sufficiency of this condition arises from \cite{se}, the fact
that it is not a necessary condition comes from the result of \cite{plb3}
where it was shown that for a solution that cannot be embedded in string
theory the proposal proved right. The asymptotic behavior of the metric of
the form prescribed by Brown and Henneaux is also a sufficient condition as
shown in \cite{st}.  The work in this paper shows that it is not a necessary
condition. The only condition that seems to be necessary is that the near
horizon geometry must be of $AdS_3$ type.  Unfortunately it makes sense to
speculate that some sort of supersymmetry can be a necessary condition
because the proposal works only for near extremal black holes and in some
sense extremality is related to supersymmetry.  There is still some hope
(raised by \cite{se} and \cite{boer}) that the required supersymmetry is lower
than the one required in \cite{sv}. Certainly some work is still needed to
clarify this issue.

Hopefully arguments similar to those presented here could be applied to the
general case of the $AdS/CFT$ correspondence of which Strominger's proposal is
a special case. In particular it seems possible to generalize the
$AdS/CFT$ correspondence to any space that needs not be of the $AdS$ type but
needs only to be dual to the $AdS$ in some sense. Furthermore, this could
eventually include string theory in asymptotically flat spaces.

\begin{center}
{\large \bf Acknowledgments}
\end{center}
It is a pleasure to thank M.Z. Iofa, Y.S. Myung and I. Sachs for
useful comments on a preliminary version of this paper. I am especially
grateful to Robert Myers for sharing some of his insights with me.

\end{document}